\documentclass[]{mn2e}
\usepackage{psfig}
\usepackage{longtable}

\def\Liso{\mbox{$L_{\rm iso}$}}
\def\Lisop{\mbox{$L_{\rm iso,p}$}}
\def\Eiso{\mbox{$E_{\rm iso}$}}
\def\Ep{\mbox{$E_{\rm p}$}}
\def\Epr{\mbox{$E_{\rm p}^\prime$}}
\def\Epri{\mbox{$<E_{\rm p}^\prime>$}}
\def\tdu{\mbox{$T_{f}$}}
\def\tdur{\mbox{$T_{f}^\prime$}}

\def\ppf{\mbox{$F_{\rm p}$}}
\def\f{\mbox{$F$}}  
\def\tbr{\mbox{$t_{\rm j}$}}
\def\nufnu{\mbox{$\nu f_\nu$}}
\def\thetaj{\mbox{$\theta_j$}}

\def\funits{\mbox{phot\ cm$^{-2}$ s$^{-1}$}}
\def\sintr{\mbox{$\sigma_{\rm intr}$}}
\def\sinto{\mbox{$\sigma_{\rm intr}^{\rm tot}$}}
\def\Lga{\mbox{$L_{\gamma}$}}
\def\thj{\mbox{$\theta_{j}$}}

\def\sw{\mbox{\textit{Swift}}}
\def\cgro{\mbox{\textit{CGRO}}}
\def\apj{\mbox{ApJ}}
\def\mnras{\mbox{MNRAS}}
\def\apjl{\mbox{ApJL}}

\newcommand{\ergcms}{erg cm$^{-2}$ s$^{-1}$}

\def\spose#1{\hbox to 0pt{#1\hss}}
\newcommand\lsim{\mathrel{\spose{\lower 3pt\hbox{$\mathchar"218$}}
     \raise 2.0pt\hbox{$\mathchar"13C$}}}
\newcommand\gsim{\mathrel{\spose{\lower 3pt\hbox{$\mathchar"218$}}
     \raise 2.0pt\hbox{$\mathchar"13E$}}}

\title[Time--resolved spectral correlations of long--duration Gamma--Ray Bursts]
{Time--resolved spectral correlations of long--duration GRBs}

\author[Firmani et al.]
{C. Firmani$^{1,2}$\thanks{E--mail: firmani@merate.mi.astro.it}, 
J.I. Cabrera$^{1}$, V. Avila--Reese$^{1}$, G. Ghisellini$^{2}$, 
G. Ghirlanda$^{2}$,
 \newauthor
L. Nava$^{2}$ and Z. Bosnjak$^{3}$\\
$^{1}$Instituto de Astronom\'{\i}a, Universidad Nacional Aut\'onoma de M\'exico
A.P. 70-264, 04510, M\'exico, D.F.\\
$^{2}$INAF--Osservatorio Astronomico di Brera, via E.Bianchi 46, I-23807 Merate, Italy\\
$^{3}$Institut d'Astrophysique de Paris, UMR 7095 CNRS -- 
Université Pierre et Marie Curie - Paris 6,\\
98bis boulevard Arago, 75014 Paris, France\\
}

\begin{document}
\pagerange{\pageref{firstpage}--\pageref{lastpage}} \pubyear{2002}

\maketitle

\label{firstpage}

 \begin{abstract}
For a sample of long Gamma--Ray Bursts (GRBs) with known redshift, we study the distribution of the 
evolutionary tracks on the rest--frame luminosity--peak energy \Liso--\Epr\ diagram. 
We are interested in exploring the extension of the `Yonetoku' correlation to any phase of 
the prompt light curve, and in verifying how the {\it high--signal} prompt duration time, 
\tdur~ in the rest frame correlates with the residuals of such correlation (Firmani et al. 2006).
For our purpose, we analyse separately two samples of time--resolved spectra corresponding
to 32 GRBs with peak fluxes $\ppf>1.8$ \funits~ from the \sw--BAT detector, and 7 bright GRBs 
from the \cgro--BATSE detector previously processed by Kaneko et al. (2006). 
After constructing the \Liso--\Epr\ diagram, we discuss the relevance of selection effects,
finding that they could affect significantly the correlation.  However, we find that these
effects are much less significant in the  \Liso \tdur--\Epr\ diagram, where the intrinsic 
scatter reduces significantly.
We apply further corrections in order to reduce the intrinsic scatter even more. 
For the sub--samples of GRBs (7 from \sw\ and 5 from \cgro) with measured jet break time, \tbr, 
we analyse the effects of correcting \Liso\ by jet collimation. 
We find that (i) the scatter around the correlation is reduced, and 
(ii) this scatter is dominated by the internal scatter of the individual evolutionary tracks. 
These results suggest that the time integrated `Amati' and `Ghirlanda' correlations 
are consequences of the time resolved features, not of selection effects,
and therefore call for a physical origin. 
We finally remark the relevance of looking inside the nature of the evolutionary tracks.

\end{abstract}


\section{Introduction}

The phenomenological study of long Gamma--Ray Bursts (GRBs), on one hand, brings the 
possibility to use these extreme cosmic events as tracers of many astronomical and 
cosmological processes, and on the other, leads to a comprehensive 
description of the underlying GRB physics (see for recent reviews M\'esz\'aros 2006;
Ghirlanda, Ghisellini \& Firmani 2006; Zhang 2007). 
The finding of correlations among spectral and energetic properties of the prompt emission 
has been the main avenue of progress in these directions. 

Amati et al. (2002) discovered a correlation between the isotropic--equivalent 
energy radiated during the whole prompt phase, \Eiso, and the peak energy of the 
rest--frame {\it time--integrated} \nufnu\ spectrum, \Epri. 
The scatter around the \Eiso--\Epri\ correlation suggests the existence of some hidden 
parameter(s). 
Ghirlanda, Ghisellini \& Lazzati (2004) have identified such parameter with the jet collimation 
angle, \thetaj, and Liang \& Zhang (2005) with the rest--frame break--time in 
the afterglow light curve, \tbr; \thetaj\ is tightly associated to \tbr. 
Both parameters are time integrated quantities of the outflow in the fireball model. 
A correlation analogous to the `Amati' one but for the peak isotropic--equivalent luminosity, 
\Lisop, has been first reported by Yonetoku et al. (2004).  
Later on, for a sample of 19 GRBs, Firmani et al. (2006) have found a strong correlation 
among \Lisop, \Epri~ and a rest--frame prompt emission {\it high--signal} GRB time duration, \tdur.   
Rossi et al. (2008) have also found a correlation among \Lisop, \Epri, and \tdur, but with different slopes 
and scatter.
More recently, Collazzi \& Schaefer (2008) confirmed the 'Firmani' correlation, though the scatter
they obtained for a larger sample is higher than in Firmani et al. (2006).  
The homogeneity and the quality of the data seems to be crucial for obtaining reliable statistical results; 
in this sense, the publicly available large database from the \sw~ is by now an invaluable source of data.

The GRB light curves are composed by a sequential superposition of many pulses, 
possibly originated by individual shock episodes. 
The spectral and temporal characteristics of these pulses are key ingredients for 
understanding the prompt emission mechanisms of GRBs (e.g., Ryde \& Petrosian 2002). 
Although there is not a simple pattern to describe the great variety of the light curve 
morphologies, some relations have been found in GRBs with light curves characterized by 
a single pulse or a few well--separated pulses. 
For example, Borgonovo \& Ryde (2001) have found a correlation between the (bolometric 
or spectral peak) energy flux, \f, and observed \Ep\ during the decaying phase of individual, 
well--defined pulses: $\f\propto \Ep^\gamma$, with a mean $\gamma\approx 2$. 
Notice that \Ep\ in this case is inferred from each time--resolved spectrum. 
Further, Liang, Dai \& Xu (2004) have tested whether or not the relation $\f\propto \Ep^2$ 
within a burst holds in a sub--sample of 2408 time--resolved spectra for 91 bright \cgro-BATSE 
GRBs presented by Preece et al. (2000; all kind of pulses are included in this sample).  
They reported that for 75\% of the bursts, the Spearman correlation coefficient is larger 
than 0.5. 
Nevertheless, even for the Borgonovo \& Ryde sample of well--defined pulses, this correlation 
has a considerable scatter, which is probably produced by the superpositions of pulses.

If the individual pulses within a burst follow on average the correlation $\f\propto \Ep^2$, 
then a natural question is 
{\it whether there exists or not a universal relationship among all GRBs in the rest frame.
}
 
Both the `Yonetoku' and the \Lisop--\Epri--\tdur\ correlations  (Firmani et al. 2006)  were 
established for GRBs with known $z$ taken from heterogeneous samples, with incomplete
available data, and by using hybrid quantities: the flux used to calculate \Lisop\ is at the peak 
of the light curve, while both \Epri\ and the bolometric correction had to be inferred 
from the time--integrated spectrum. 
The primary data publicly available from the \sw\ satellite (Gehrels et al. 2004) offers now the 
possibility to study the mentioned correlations in the rest--frame and with temporal resolution. 
Unfortunately, the narrow spectral range of the \sw\ BAT detector limits and makes difficult the 
spectral analysis.

By means of time--resolved spectral analysis for a sample of 32 \sw\ and 7 bright \cgro\ long 
GRBs, we will show here that, 
when multiplying \Liso\ by \tdur, the evolutionary tracks of different GRBs form a well defined 
correlations, with a small scatter. 
In \S 2 we develop a heuristic framework to introduce such a connection.  
The GRB samples used and the spectral analysis are presented in \S 3. 
The construction of correlations with the prompt time--resolved data, 
the analysis of the selection effects, and a discussion of the results 
are presented in \S 4.1. 
In \S 4.2, we present our attempt to further reduce the scatter around the 
correlations by introducing the jet collimation correction. 
We show that most of the remaining scatter in the overall correlations is due 
to the individual scatter of each evolutionary track. 
In \S 5 we present our conclusions.

\begin{figure}
\centerline{
\psfig{file=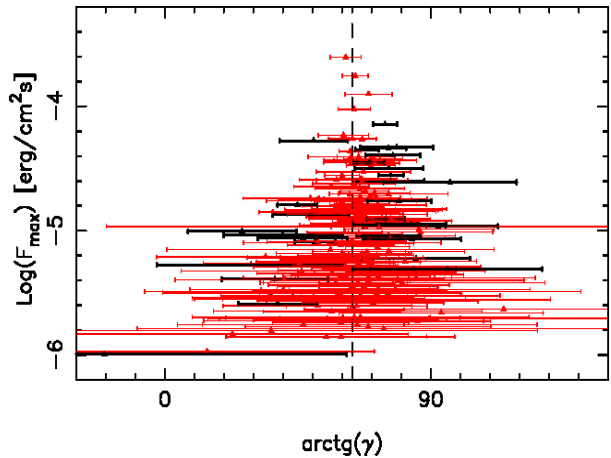,width=8cm,height=6cm} }
\caption
{Maximum bolometric flux of a given event versus $\arctan(\gamma)$ for 208  \cgro\ GRBs 
with 5 or more time--resolved spectra (from the Kaneko et al. 2006 sample). 
The quantity $\gamma$ is the slope of the best linear fit to the $F$--\Epr\ evolutionary
tracks. The $\arctan$ function is used in order the error bars ($3 \sigma$) looked symmetric. 
The vertical dashed line indicates $\arctan 2=63.4$ degrees.
Thin red error bars show the events with $\gamma= 2$ within $3 \sigma$, while thick black
error bars indicate the few events incompatible with  $\gamma= 2$ at the $3 \sigma$ level.
There is not any systematical trend of $F_{\rm max}$ with $\gamma$, however, the uncertainty 
and the spread of $\gamma$ around 2 increases as $F_{\rm max}$ decreases.
}
\label{Fgamm}
\end{figure}

\section{Motivations}

Our purpose in this paper is to study, in the rest frame, evolutionary features of the prompt 
$\gamma-$ray spectra and their connection with other GRB properties.
Before discussing the data, we develop an heuristic reasoning aimed to show a potential 
connection between the phenomenological time--integrated and time--resolved correlations 
mentioned in the Introduction.

We start from the fact that the prompt evolution of a given burst can be characterised, at least 
above some flux and/or some peak energy limit, by the dependence $\f \propto \Ep^\gamma$, 
where \f\ and \Ep\ are defined at any given time. 
The power index $\gamma$ as well as the scatter of the evolutionary correlation could change 
from burst to burst. 
Previous analysis showed that the mean value of $\gamma$ is around 2 with some dispersion 
(Borgonovo \& Ryde 2001; Liang et al. 2004). 
In the rest frame, the burst evolutionary track $\f \propto \Ep^\gamma$ is equivalent to the correlation 
\begin{equation}
\Liso=K \Epr^\gamma.
\label{LEC}
\end{equation}
In the logarithmic plane \Epr--\Liso\ the evolutionary track described by Eq. (\ref{LEC}) 
identifies a straight strip with a slope $\gamma$, a zero point, and a specific sector on the strip. 

Let us first assume $\gamma=2$.
Note that $K$ in this case is independent of the Doppler--boosting factor 
$D=1/[\Gamma(1-\beta \mu)]$, where $\Gamma$, $\beta c$, and  $\mu$
are the Lorentz factor, the velocity, and the cosine of the viewing angle, respectively.
By integrating in time, the evolutionary track reduces to 
\begin{equation}
\Eiso = KT^\prime <\Epr^2>,
\label{amati}
\end{equation}
where $T^\prime$ is a rest--frame time scale of the energy emission.
Now, if observations give a correlation between \Eiso\ and $\Epri^2$ among different GRBs,
(the `Amati' relation), then $KT^\prime=k$ becomes approximately a constant independent of each GRB.
The conclusion is that scaling $K \propto T^{\prime-1}$, the scatter of the correlation 
given by Eq. (\ref{LEC}) for a sample of GRBs is less than the scatter around the \Liso--\Epri\  
correlation for the same sample.
The fact that $\gamma = 2$ for all the GRBs simplifies the previous reasoning: since the 
evolutionary track given by Eq. (\ref{LEC}) and the `Amati' relation show the same slope, then only 
the zero point of the strip is involved in the scatter around the `Amati' relation.  
The parallel strips superimpose and thus reduce their occupation region on the plane. 

If $\gamma \neq 2$, the situation becomes more intriguing. 
Now the evolutionary tracks are not aligned with the `Amati' relation 
and so it is more difficult for them to reduce their occupation region on the  
\Epr--\Liso\ diagram. 
Nevertheless, if not only the zero points of the strips, but even the occupation 
sectors of the evolutionary tracks within the strips {\it conspire} in order to 
agree with the `Amati' relation, then a reduction of their occupation region on 
the plane (reduction of the scatter) may  be reached. 
This inference allows to hope that $T^\prime$ plays a role of a hidden parameter related 
to the scatter of the correlation Eq. (\ref{LEC}). 

This heuristic reasoning has actually inspired the present work. 
Definitively, it is of great interest to verify with high quality prompt 
time--resolved data whether or not the `Firmani'--like time--resolved relation 
$\Epr$ vs $\Liso T^\prime$ is obeyed, and to study the scatter around it.  
If this relation is obeyed, then a corollary is that it establishes 
a link between the `Amati' $\Eiso-\Epri$  and the `Yonetoku' $\Lisop-\Epri$ 
relations and, even more importantly, a robust physical ingredient has to 
exist behind such relation in the sense that at each instant of the prompt 
the GRB {\it is aware} of the duration of the entire process.

Before presenting results from our spectral analysis on \sw\ and \cgro\ GRBs with known
redshift, we explore the large set of time--resolved spectra from the \cgro\ data analysed 
by Kaneko et al. (2006). Since for the great majority of the bursts in this sample the 
redshift is unknown, we are able to infer only the values of the slope $\gamma$
of the individual correlations $F\propto\Ep^\gamma$. The slope should remain the same
for the corresponding individual rest--frame correlations $\Liso$ vs $\Epr$.

The Kaneko et al. (2006) sample contains 350 bright GRBs and 7427 time--resolved spectra. 
Each time--resolved spectrum was fitted to a Band (Band et al. 1993) function.
We consider only time--resolved spectra with Band model parameters $\alpha > -2$; 
$\beta<-2$ and 50$<$\Ep$<$1800 keV (note that \Ep\ is in the observer
frame) with an uncertainty better than 50\% and a time step shorter than 4 s. 
We assume a bolometric correction given by the Band energy distribution. 
Finally, we take into account only GRBs with at least 5 time--resolved spectra. 
Such conditions reduce the sample to 208 GRBs. 

\begin{figure*}
\vspace{11cm}
\hspace{13cm}
\includegraphics{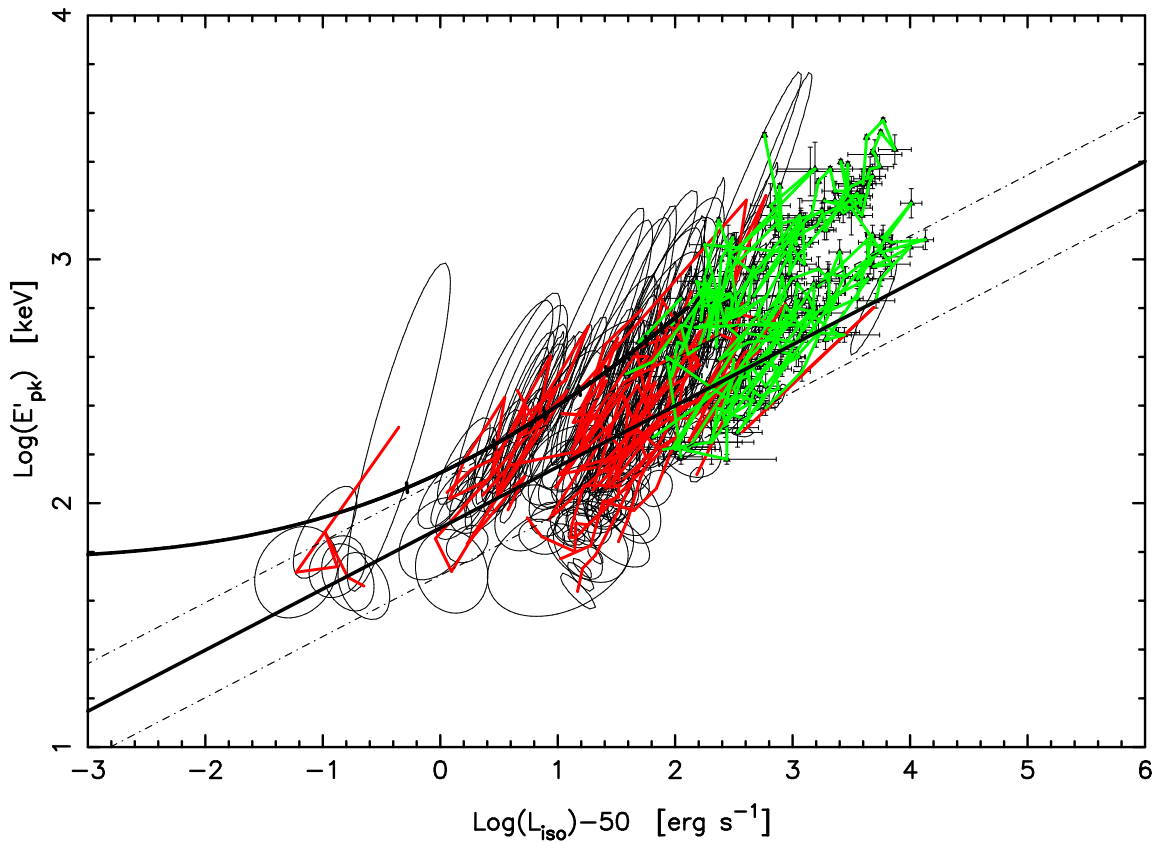}
\caption{$\Epr$ vs $\Liso$ for the time--resolved spectra corresponding to 
32 \sw\ GRBs ($68 \%$ CL ellipses) and 7 bright \cgro\ GRBs ($1\sigma$ error bars). 
The solid lines connect the data--points of each individual GRB (evolutionary tracks); 
red colour is for the \sw\ data and green colour for the \cgro\ data.
The best linear fit to the \sw\ data is shown with the solid right line, while 
the $\pm 1\sigma$ intrinsic scatter is indicated with the parallel dot--dashed lines. 
The upper solid curve corresponds to a limiting observed ($z=0$) flux of 
$10^{-8}$\ergcms\ and $\Ep\simeq 100$ keV. The ticks along the curve 
indicate different $z'$s: 1, 2,...,9, from left to right.}
\label{LESB}
\end{figure*}

For this sample, only 16\% of the GRB tracks show a disagreement with $\gamma = 2$, while the 
other 84\% are compatibles with $\gamma = 2$ at the $3\sigma$ level. A similar result, but for
less events than here, has been reported by Liang et al. (2004; see their figure 2 for the $F-\Ep$ 
tracks corresponding to some bursts).  
We plot  in Fig. \ref{Fgamm} the maximum bolometric flux, $F_{\rm max}$, of a given event versus 
$\arctan(\gamma)$; $\arctan$ is used in order the error bars (at 3$\sigma$) looked symmetric. 
It is clearly seen that only a small fraction of the events (thick black error bars) have
values of $\gamma$ different from 2 within $3 \sigma$ (vertical dashed line shows $\arctan 2=63.4$ degrees);
the majority of them (thin red error bars) show $\gamma$ compatible with 2 within the $3 \sigma$ level. 
In  Fig. \ref{Fgamm}  it is also seen that the correlation slope $\gamma$ is not biased systematically 
by $F_{\rm max}$, though its uncertainty and the spread around 2 increse as $F_{\rm max}$ decreases.
The result shown above opens the question of identifying the physical mechanisms that determine the 
value of $\gamma$ as well as the intrinsic scatter of each evolutionary track. 
The fact that a large fraction of GRBs shows $\gamma \simeq 2$, 
$(i)$ supports the consistency of our research concerning a universal correlation, 
$(ii)$ provides a natural GRB identification criterion to eventually optimise such correlation, and 
$(iii)$ allows to study the behaviour of those GRBs with $\gamma \neq 2$.

\begin{table*}\centering
\begin{tabular}{lccccccc}
\hline\hline
name    &      z      &      \tdu         &          \tbr          & $\rm{Lg} <\Ep^{\prime}>$ & $\rm{Lg}\Eiso$ &   N      \\
        &             &         (sec)            &          (days)            &       (keV)
            &   ($10^{50}$ erg) &          \\        
\hline
050315   &  1.949 &   15.00[0.50] &                    &                    &                    &   \ 1  \\ 
050318   &  1.440 &   \ 4.00[0.10] &                    &   2.45[0.04] &   2.07[0.11] &  \ 2  \\ 
050401   &  2.900 &   \ 6.20[0.10] &   1.50[0.50] &   3.45[0.23] &   3.51[0.14] &  \ 3  \\ 
050505   &  4.270 &   11.80[0.10] &                    &   3.54[0.16] &   3.13[0.12] &  \ 2  \\ 
050525   &  0.606 &   \ 3.10[0.02] &   0.30[0.10] &   2.33[0.01] &   2.34[0.01] &  19  \\ 
050603   &  2.821 &   \ 2.10[0.10] &                     &   3.68[0.30] &   3.55[0.22] &  \ 3  \\ 
050820A &  2.612 &   12.00[0.40] & 15.00[8.00] &   3.12[0.09] &   3.99[0.03] &  \ 2  \\ 
050922C &  2.198 &   \ 1.44[0.02] &                     &   3.36[0.22] &   2.63[0.14] &  \ 6  \\ 
051109   &  2.346 &   \ 6.90[0.20] &                     &   2.97[0.16] &   2.80[0.19] &  \ 1  \\ 
051111   &  1.549 &   11.60[0.10] &                     &   3.36[0.34] &   2.80[0.22] &  \ 1  \\ 
060206   &  4.048 &   \ 2.70[0.10] &                     &   3.29[0.06] &   2.59[0.06] &  \ 5  \\ 
060210   &  3.910 &   35.00[3.00] &   2.20[0.90] &   3.45[0.14] &   3.52[0.10] &  \ 4  \\ 
060418   &  1.490 &   16.20[0.30] &                     &                    &                     &  \ 3  \\ 
060526   &  3.221 &   12.30[0.30] &   2.80[0.30] &   2.02[0.09] &   2.41[0.05] &  \ 1  \\ 
060614   &  0.125 &   28.70[0.10] &   1.38[0.04] &   1.74[0.36] &   1.40[0.17] &  \ 6  \\ 
060904B &  0.703 &   10.00[0.30] &                    &   2.36[0.13] &   1.44[0.13] &  \ 5  \\ 
060906   &  3.685 &   11.50[0.20] &                    &   2.99[0.09] &   3.16[0.14] &  \ 1  \\ 
060927   &  5.600 &   \ 4.10[0.20] &                    &   3.50[0.06] &   2.92[0.06] &  \ 4  \\ 
061007   &  1.262 &   18.40[0.08] &                    &                     &                    &  \ 5  \\ 
061121   &  1.314 &   \ 6.20[0.05] &                    &                     &                    &  \ 3  \\ 
070318   &  0.836 &   12.00[0.70] &                    &                     &                    &  \ 9  \\ 
070508   &  0.820 &   \ 6.10[0.05] &                    &                     &                    &  20  \\ 
070521   &  0.553 &   10.30[0.20] &                    &   2.66[0.10] &   2.11[0.07] &  22  \\ 
070810A &  2.170 &   \ 3.30[0.10] &                    &   2.58[0.07] &   2.14[0.15] &  \ 5  \\ 
071003   &  1.100 &   12.80[0.50] &                    &                    &                     &  \ 9  \\ 
071010B &  0.947 &   \ 4.42[0.05] &                    &   2.25[0.04] &   2.29[0.09] &  14  \\ 
071117   &  1.331 &   \ 1.50[0.10] &                    &   2.99[0.21] &   2.21[0.12] &  \ 7  \\ 
080319B &  0.937 &   21.11[0.07] &   9.00[2.00] &   3.12[0.01] &   4.12[0.01] &  \ 2  \\ 
080319C &  1.950 &   \ 5.16[0.08] &                    &   3.15[0.15] &   2.63[0.11] &  \ 5  \\ 
080411   &  1.030 &   \ 6.55[0.02] &                    &                    &                     &  21  \\ 
080413A &  2.433 &   \ 6.37[0.03] &                    &                    &                     &  \ 9  \\ 
080413B &  1.100 &   \ 1.15[0.01] &                    &   2.59[0.12] &   2.19[0.10] &  \ 7  \\ 
\hline\hline
\end{tabular}
\caption{Sample of \sw--BAT long GRBs with known $z$ and useful time--resolved spectra.
The prompt observer emission time duration \tdu\ was calculated by us (see text), while
the break time \tbr\ was taken from the literature compilation by Ghirlanda et al. (2007);
for GRB080319B, \tbr\ was taken from Kann, Schulze \& Updike (2008). 
The time--integrated rest--frame peak energy and isotropic--equivalent energy were taken 
from Cabrera et al. (2007) or calculated by us using the CPL model. 
$N$ is the number of useful time--resolved spectra.
}
\end{table*}

\begin{table*}\centering
\begin{tabular}{lccccccc}
\hline\hline
name    &      z      &        \tdu         &         \tbr         & $\rm{Lg} <\Ep^{\prime}>$ & $\rm{Lg}\Eiso$ &   N      \\
        &             &         (sec)            &          (days)            &       (keV)
            &   ($10^{50}$ erg) &          \\

\hline
970828  &  0.957 &   11.00[2.00] &   2.20[0.40] &   3.06[0.09] &   3.47[0.05] &  35  \\ 
980703  &  0.966 &   18.80[0.30] &   3.40[0.50] &   2.99[0.09] &   2.84[0.05] &  \ 2  \\ 
990123  &  1.600 &   19.50[0.10] &   2.00[0.50] &   3.72[0.03] &   4.38[0.05] &  84  \\ 
990506  &  1.307 &   15.30[0.20] &                    &                     &                    &  80  \\ 
990510  &  1.619 &   \ 5.50[0.10] &   1.60[0.20] &   3.05[0.04] &   3.25[0.05] &  20  \\ 
991216  &  1.020 &   \ 4.30[0.05] &   1.20[0.40] &   3.12[0.09] &   3.83[0.05] &  54  \\ 
000131  &  4.500 &   \ 2.00[0.50] &                    &                     &                    &  12  \\ 
\hline\hline
\end{tabular}
\caption{Sample of \cgro--BATSE long GRBs with known $z$ and useful time--resolved spectra
calculated by Kaneko et al. (2006). 
In this case, the Band model was used for both the time--resolved and time--integrated quantities. 
See the caption of Table 1 for more details.}
\end{table*}

\section{Sample selection and spectral analysis}

The time--resolved spectral analysis requires data with a high enough signal--to--noise 
ratio in order to ensure a reasonable determination of the spectral parameters.
We select here 32 long GRBs from the whole \sw--BAT sample of GRBs with known 
redshift (until February 2008) and with a peak flux, \ppf, greater than $1.8 \  \funits$.
Because of the \sw--BAT narrow effective spectral range (15--150 keV), a careful
statistical handling of the correlated uncertainties in the spectral parameters is 
mandatory (Cabrera et al. 2007). 
Even so, a significant fraction of the observed GRBs and of the time--resolved spectra 
within a given GRB, had to be discarded because the peak in their \nufnu\ spectra lies 
out of the BAT limit or because the signal is too low.
At the end, we have 32 \sw\ usable bursts with 207 time--resolved spectra.

The photon model adopted to fit the \sw\ time--resolved spectra is the cut--off power law 
(CPL), which has three parameters. 
The fits are carried out with the {\it heasoft} package XPSEC
\footnote{http://heasarc.gsfc.nasa.gov/docs/software/lheasoft/download.html}.
The time--resolved spectra selected for the analysis are chosen to be shorter than 
the light curve variation time scales, but large enough as to ensure 
acceptable confidence levels (CLs) for the photon index and \Ep.
We reject the prompt time--resolved spectra where \Ep\ can not be determined
or the uncertainty on \Ep\ and/or on the photon index is too large due to the 
scarce number of counts.
With these constraints, we obtained 207 usable time slices for our 32 \sw\ GRBs. 
Table 1 reports the basic information concerning this sample: the name, the redshift, 
the observer--frame \tdu\ and jet break time \tbr, and finally the rest--frame time--integrated 
\Epr\ and total \Eiso .
\tdu\ is the time spanned in the observer frame by the brighest $50 \%$ of the total counts 
above background (Reichart et al. 2001) for the light curve on the observed energy 
range $17-100$ keV 
\footnote{We measure \tdu~ as in Reichart et al. (2001), but instead of using the 
duration of the brightest bins in the light curve that enclose $f=45\%$ of the total counts, 
we have used $f=50\%$. None of the results presented here changes by such redefinition. 
}.
A more detailed description of the temporal binning and selection criteria will be presented 
elsewhere.  
\Liso\ is calculated from the CPL spectral parameters within the energy range
1--10000 keV at rest. 
The correlated errors of the spectral parameters are adequately propagated in order to 
obtain the corresponding correlated errors (CL ellipses) of \Liso\ and \Epr\ (see for details 
Cabrera et al. 2007). 

With the aim to compare it with a completely different sample, along with the $32$ GRBs observed 
by \sw, we have included in our considerations $7$ bright \cgro--BATSE GRBs with known redshifts
reported in Table 2. 
Now \tdu\ is measured on light curves in the $50-300$ keV observed energy range.
For these 7 GRBs we have 287 useful time--resolved spectra available from Kaneko et al. (2006).
In the  case of the \cgro\ sample, the spectral information tends to be of much better quality 
and the time--resolved spectra can be fitted with the more general four--parameter Band model. 

The \sw--BAT and \cgro--BATSE samples studied here are different in many aspects.
We remark two of them. 
The first one is due to the spectral model, CPL for \sw\ and Band for \cgro. 
Based on Cabrera et al. (2007), a rough estimate of this difference may be obtained adopting 
for the \sw\ spectral fitting a Band model with $\beta$ frozen to $-2.3$. 
The result shows an increase in \Liso\ by a factor $1.6$ for a given \Epr. 
The second difference is the light--curve energy range in which \tdu is estimated.
For \sw\ we use $17-100$ keV, while for \cgro, we use $50-300$ keV.
Taking into account the inverse dependency of \tdur\ on the energy at the power $0.4$ given 
by Reichart et al. (2001), the factor $3$ between \sw\ and \cgro\ energy ranges leads to a 
\tdur\ for the \sw\ sample roughly $3^{0.4} = 1.55$ larger than \tdur\ for the \cgro\ sample. 
Curiously enough, both effects leave the product $\Liso\times\tdur$ roughly invariant.
In spite of this coincidence, we have decided to handle each one of the samples separately, 
and only afterwards we will take care to compare the results of our analysis that 
result invariant with respect to the systematic differences mentioned above.

\section{Correlation analysis}

\subsection{Prompt $\gamma$--ray emission correlations}

Figure \ref{LESB} shows in the \Liso--\Epr\ diagram the time--resolved spectral data for the \sw\ 
sample (ellipses) and for the \cgro\ sample (error bars). The ellipses correspond to the 
$68\%$ CL error regions calculated taking into account the error covariance matrix 
(see Cabrera et al. 2007), while the orthogonal error bars correspond to the standard deviations. 
No correction for the differences in the spectral model has been applied (see \S 2). 
The scatter in the plot is rather large. 
The red lines show the evolutionary tracks for  the \sw\ sample.
The brightest (highest S/N ratios) bursts show $\gamma\simeq 2$ while the fainter bursts show 
values between $1$ and $4$. It is not possible to obtain more continuous and detailed tracks
because of relevant sections of the light curves that have time--resolved spectra with $\Ep$ lying 
outside the BAT spectral sensitivity. The green lines show the evolutionary tracks 
for the \cgro\ sample. Four events show $\gamma=2$ at 1$\sigma$, two (990123 and 990510) at 
2$\sigma$, and one (990506) at 3$\sigma$.
We have shown in \S 2 that even for the much larger sample of \cgro\ GRBs without redshift 
determination (from Kaneko et al. 2006), indeed $\gamma\approx 2$ for most of the events.

The bottom right part in the \Liso--\Epr\ diagram is physically empty; here, the selection 
effects do not apply.  Therefore, a real upper limits $\Liso$ for each $\Epr$ does exist. 
On the contrary, in the top left part of the diagram, selection effects could be present. 
The upper continuous curve plotted in Fig. \ref{LESB} corresponds to a limiting hypothetical 
observed flux of $10^{-8}$\ergcms\ between 15--150 keV and an observed peak energy 
$\simeq 100$ keV. This curve gives a rough idea of the sensitivity limit of the \sw--BAT 
instrument.  Therefore, the present data do not allow to identify a specific boundary here 
because of  the limiting fluxes characterizing the samples. 
Thus, taking care of eventual selection effects in this region of the diagram, a kind of `Yonetoku' 
correlation may be extended to the time--resolved features and may include a conspicuous fraction 
of the prompt evolution. This correlation might be reflecting an intrinsic local physical process of 
the GRB emission mechanism.  

 Along the constant flux/\Ep\ curve plotted in Fig.  \ref{LESB}, we show with ticks the values that 
would have \Epr\ and \Liso\ at different redshifts (integer $z$ from 1 to 9). 
This result shows that this correlation would not be useful to infer pseudo--redshifts for GRBs 
with non determined redshift. 
A similar conclusion will apply for the \Epr--\Liso\tdur\ correlation presented below. 

\begin{figure}
\centerline{
\psfig{file=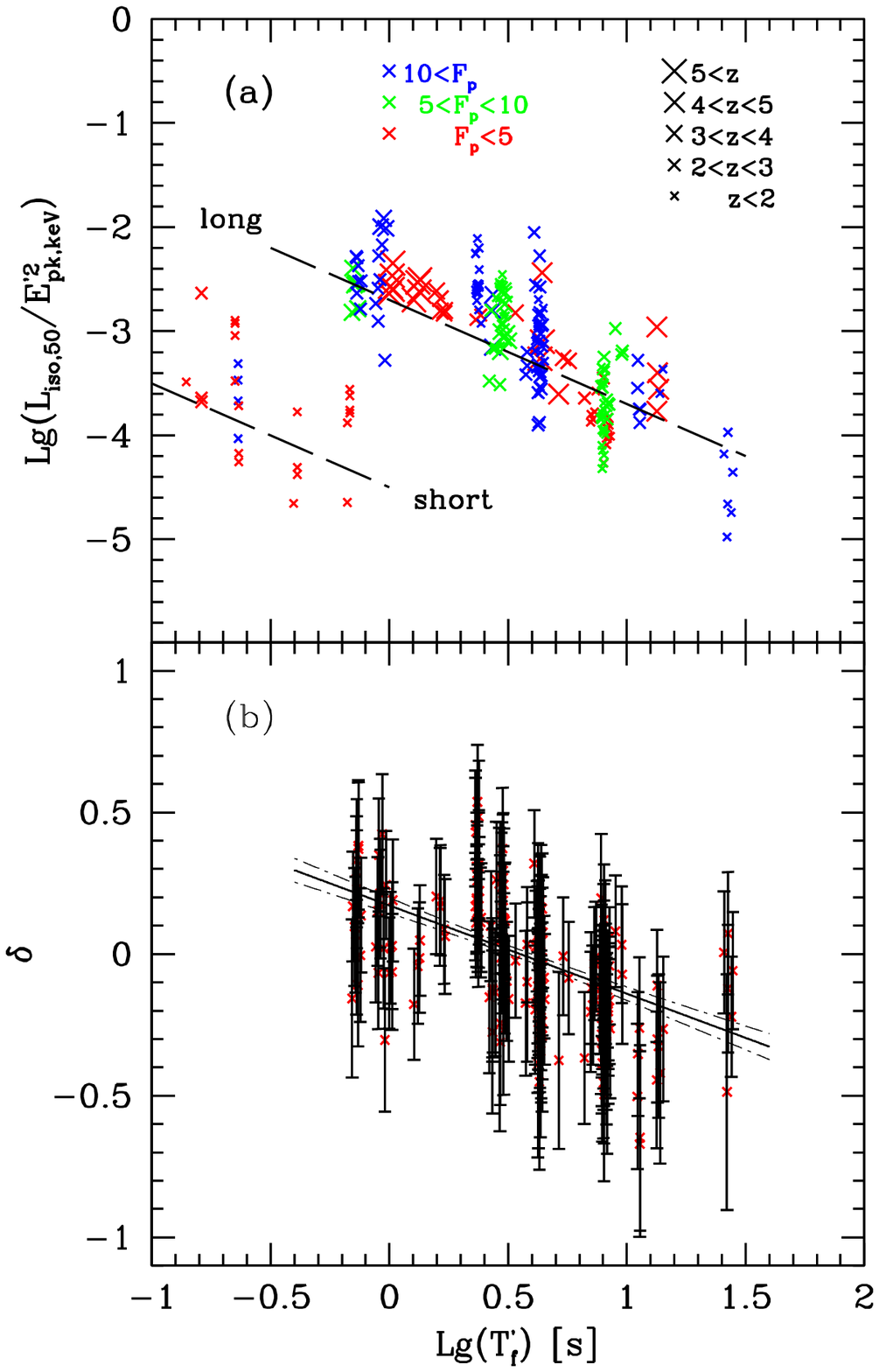,width=14cm,height=14cm} }
\caption{
{\it Panel (a)}: $\Liso/\Epr^2$ vs $\tdur$ for each burst from the same \sw\
time--resolved spectral data shown in Fig. \ref{LESB} and for 9 short \sw\ GRBs. 
The size represents the redshift while the colour represents the peak flux in \funits. 
The dashed lines correspond to a $-1$ slope.
{\it Panel (b):} Residuals $\delta$ of the correlation \Liso-\Epr\ shown in Fig. \ref{LESB}
and their corresponding uncertainties vs $\tdur$. 
The best fit slope $-0.30 \pm 0.04$ leads to a scatter reduction
term proportional to $\tdur^{1.25 \pm 0.20}$.
}
\label{LEvsT}
\end{figure}

The thick straight solid line in Fig. \ref{LESB} is the best linear fit to the logarithmic data 
performed by a minimum $\chi^2$ method taking into account the residual and its uncertainty 
on the orthogonal direction of the straight line. In Fig. \ref{LESB} (as well as in the following 
related Figures), in order to avoid overcrowding, the best fit is plotted only for the 
\sw\ sample. The fit gives a line described by: 
\begin{equation}
{\rm Log}\left(\frac{\Epr}{\rm keV}\right) =  
a \ {\rm Log}\left( \frac{\Liso}{10^b \ {\rm erg \ s^{-1}}}\right)+c,
\label{corr} 
\end{equation}
The corresponding coefficients of the fits to both the \sw\ and \cgro\ samples are given in
Tables 3 and 4, respectively (relation LE). 
Notice that $b$ is not a parameter of the fit,
but is the logarithm of \Liso\ in the barycentre of the data--points.
The standard errors were computed in the barycentre frame of the data--points in order 
to minimise any correlation between them. The average correlation slope of both samples is 
$0.30 \pm 0.05$, and the difference in the intercepts at $\Epr = 100$ keV between the \cgro\ 
and \sw\ best fits is $0.09 \pm 0.11$. 
We have also estimated the {\it intrinsic scatter} \sintr\ of the correlation 
by adding it iteratively in quadrature to the error in the $y$ axis (Log\Epr) and by requiring 
that the reduced $\chi^2_r$ be equal to 1. This method has been tested by Novak 
et al. (2006) and it gives values similar to those obtained by the 
fitting method presented in D'Agostini (2005).  The intrinsic scatters (standard
deviations) around the $\Epr$ vs $\Liso$ correlation for both samples are given in the 
same Tables 3 and 4. The average value of the \sintr\ is $0.21 \pm 0.01$. The dot--dashed 
lines in Fig. \ref{LESB} show the intrinsic scatter around the \sw\ sample correlation.

\begin{table*}\centering
\begin{tabular}{lccccc}
\hline\hline
relation  &a              &b     &c                 &  $\sintr$             &  $ \sinto $   \\
\hline
LE   &  0.25$\pm$0.02  &  51.48  &  2.27$\pm$0.02   &    0.195$\pm$0.013    &  ...              \\
LTE  &  0.39$\pm$0.02  &  52.17  &  2.27$\pm$0.01   &    0.137$\pm$0.010    &  ...              \\
LTHE &  0.49$\pm$0.03  &  49.48  &  2.11$\pm$0.02   &    0.093$\pm$0.020    &  0.084$\pm$0.007   \\
LTWE &  0.66$\pm$0.05  &  49.42  &  2.14$\pm$0.02   &    0.105$\pm$0.020    &  0.097$\pm$0.008    \\
\hline\hline
\end{tabular}
\caption{Best--fit slopes $a$ and zero points $c$ of the different logarithmic linear relations 
studied in this paper for the \sw\ sample. 
The parameter $b$ defines the absisa of the barycentre ($10^b$, in cgs units)  of the given correlation 
while \sintr\ is the intrinsic scatter. 
For the correlations which include the collimation angle, \sinto\ is an estimate of the scatter contribution 
due to the internal dispersion of the evolutionary tracks.
}
\end{table*}

\begin{table*}\centering
\begin{tabular}{lccccc}
\hline\hline
 relation   & a         &b        &  c               &    $\sintr$            &  $ \sinto $   \\
\hline
LE    &  0.36$\pm$0.03  &  52.91  &  2.87$\pm$0.01   &    0.217$\pm$0.010     & ...                     \\
LTE   &  0.42$\pm$0.02  &  53.72  &  2.88$\pm$0.01   &    0.174$\pm$0.008     & ...                     \\
LTHE  &  0.50$\pm$0.02  &  51.23  &  2.94$\pm$0.01   &    0.104$\pm$0.007     & 0.130$\pm$0.007         \\
LTWE  &  0.53$\pm$0.02  &  50.67  &  2.93$\pm$0.01   &    0.131$\pm$0.008     & 0.130$\pm$0.007          \\
\hline\hline
\end{tabular}
\caption{Same as in Table 3 but for the \cgro\ sample.}
\end{table*}
 
Taking into account the result by Firmani et al. (2006), our next step is to explore a 
possible correlation of \Liso\ and \Epr\ with the high--signal emission
rest frame time \tdur. 
For calculating \tdur, we have taken into account the corrections
due to the cosmological time dilation and the narrowing of the light curve's temporal
substructure at higher energies (Fenimore et al. 1995). 
Following Reichart et al. (2001), this last correction on \tdu, calculated fixing the energy band
on the rest frame, goes as $(1+z)^e$ with $e \simeq 0.4$. 
Then applying both corrections we obtain $\tdur = \tdu (1+z)^{-1+e}$
Actually the value of $e$ is not well constrained. 
For a large BATSE database, Zhang et al. (2007) found that the standard deviation of the 
distribution of $e$ (modeled as Gaussian) is $0.51$ with a median value of $0.39$; the distribution 
is skewed to smaller values.  

For the \sw\ sample we find that \Liso\ roughly tends to be $\propto (\tdur)^{-1}$ with a large 
scatter, while \Epr\ does not show any significant correlation with \tdur. 
Now, if we plot $\Liso/\Epr^m$ from each time--resolved spectrum  vs $\tdur$, then the scatter reduces 
for $m \sim 2$ and a clear trend with slope $-1$ (upper dashed line) is observed (panel (a) of 
Fig. \ref{LEvsT}). 
This result implies that the residuals in the \Liso--\Epr\ diagram (Fig. \ref{LESB}) correlates with \tdur.
 
It is interesting to show in panel (a) of Fig. \ref{LEvsT} the position of some short 
GRBs with known redshift. We have carried out time resolved spectral analysis for $9$  
\sw\ short GRBs. As seen in Fig. \ref{LEvsT}, these data are naturally segregated in the 
diagram, having significantly smaller $\Liso / \Epr^2$ values than those of the long GRBs. 
    
With the aim of checking for possible selection effects behind the correlation
presented in panel (a) of Fig. \ref{LEvsT}, we plot the data with the color code
representing the observed peak photon flux (\ppf) range, and with the cross
size indicating the redshift range.  
Our first test has to do with the influence of \ppf\ on the $\Liso / \Epr^2$ vs $\tdur$ diagram. 
We have divided the sample into three sub-samples: $1.8 <\ppf \le 5$ (red), $5 < \ppf \le 10$
(green), and $\ppf > 10$ (blue) (the units are \funits).  
As seen in Fig. \ref{LEvsT}, the correlation is remarkably insensitive to the peak flux. 
In other words the sensitivity limit does not influence the correlation; it only limits the number 
of objects on it. 
Our second test refers to a possible bias with $z$. 
According to the same panel of Fig.  \ref{LEvsT}, it is evident that the data plotted in the 
\tdur--$\Liso/\Epr^2$ diagram are not appreciably biased by $z$. 
A third test concerns a possible selection effect on the GRB duration. 
By plotting $\ppf$ vs $\tdur$ for all the \sw\ GRBs with known $z$ (even those with $\ppf<1.8$
\funits, which were not used in our analysis here), we have seen that the
limit $\ppf > 1.8$ is well above any biasing limit for \tdur.

Finally, while the upper right side of the diagram in panel (a) of Fig. \ref{LEvsT} involves fluxes 
that are not biased by selection effects, the presence of short GRBs on the bottom left side of 
the same diagram gives a further evidence against any selection effect here. 
Unfortunately any further exhaustive analysis on the distribution of \Ep\ is impossible due to the 
limited BAT spectral capability. 
We conclude that the correlation $\Liso/\Epr^2$ vs $\tdur$ is reasonably free from selection effects.

Panel (b) of Fig. \ref{LEvsT} shows for the \sw\ sample how much the residuals of the \Liso--\Epr\ 
diagram of Fig. \ref{LESB} correlate with \tdur. 
Here $\delta$ are the orthogonal residuals of the \Liso--\Epr\ best fit (positive $\delta$ correspond to 
high \Liso), its standard deviation being calculated in the same direction.
The best fit slope of the $\delta$ vs $\tdur$ correlation is $-0.30 \pm 0.04$. Combining this result 
with the \Liso--\Epr\  best fit slope we obtain that the correlation $\Epr$ vs $\Liso \tdur^p$ reaches 
its minimum scatter for $p = 1.28 \pm 0.20$. 
A similar analysis on the \cgro\ sample leads to $p = 1.12 \pm 0.25$.
Given the relevance of this result we have made use of other more sofisticated methods based on 
multilinear analysis obtaining in each case $p = 1.25 \pm 0.20$ and $p = 1.00 \pm 0.20$,
respectively.
This result implies that the emission time $\tdur$ of the events on the diagram of Fig. \ref{LESB} 
increases along the orthogonal direction of the best fit straight line when one goes from the 
low \Epr\ -- high \Liso\ to the high \Epr -- low \Liso.
Therefore, as it has been found in Firmani et al. (2006) for a different sample, \tdur\  reduces the scatter 
around the correlation $\Epr$ vs $\Liso$. Later on we will estimate the reduction of such scatter. 
This result is particularly intriguing because it reveals how instantaneous features such as 
\Liso\ and \Epr\ are actually regulated according to the overall duration of the burst.

\begin{figure*}
\vspace{11cm}
\hspace{13cm}
\includegraphics{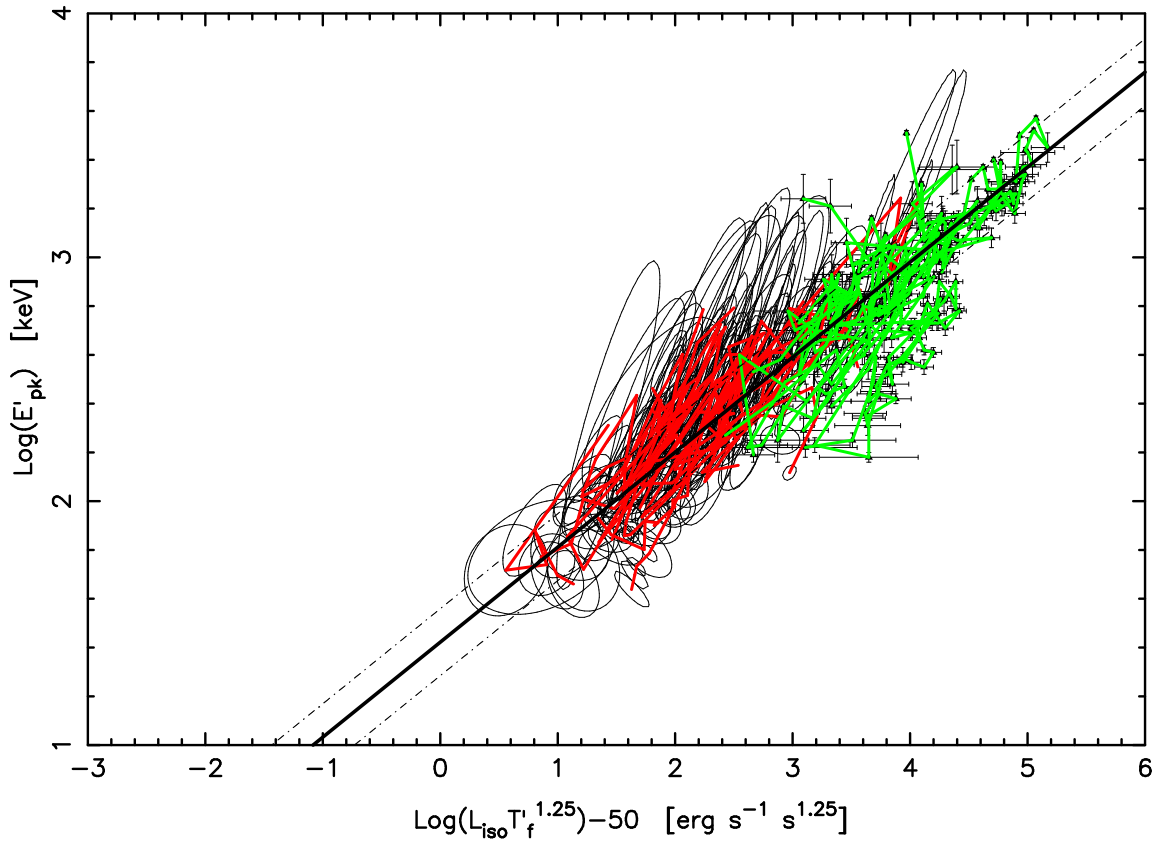}
\caption{
$\Epr$ vs $\Liso \tdur$ for the same time--resolved spectral data 
shown in Fig. \ref{LESB}. The symbol and line codes are as in Fig. \ref{LESB}.
The best--fit line and the $\pm 1\sigma$ intrinsic scatter (solid and 
dot--dashed lines, respectively) are shown only for the \sw\ sample. 
}
\label{LTESB}
\end{figure*}

We fit the data with the line
\begin{equation}
{\rm Log}\left(\frac{\Epr}{\rm keV}\right) =  
a {\rm Log}\left( \frac{\Liso \tdur^p}{10^b{\rm erg\ s^{-1}\ s^p}}\right)+c,
\label{corrt} 
\end{equation}
where we adopt $p = 1.25$ and $p = 1$ for the \sw\ and  \cgro\ samples, respectively.
The best linear fit parameters to Eq. (\ref{corrt}) for the \sw\ and  \cgro\ samples are given 
in Tables 3 and 4, respectively (relation LTE).
The average of both slopes is $0.40 \pm 0.02$, while the difference in the intercepts at 
$\Epr=100$ keV between the \cgro\ and \sw\ best fits is $0.15 \pm 0.07$.
The average intrinsic scatter is $\sintr = 0.15 \pm 0.02$. A remarkable decrease on the 
scatter from the $\Epr$ vs $\Liso$ to the $\Epr$ vs $\Liso \tdur^p$ correlation is evident.
Figure \ref{LTESB} shows $\Epr$ vs $\Liso \tdur^p$ for the same samples plotted in 
Fig. \ref{LESB}. The best fit line, as in Fig.  \ref{LESB}, refers only to the \sw\ 
sample, and the corresponding scatter is represented by the dot--dashed lines.
Our basic considerations will not change appreciably assuming $p = 1$ even for the \sw\ 
sample. In fact, the internal scatter changes from $0.137$ to $0.142$ with a standard 
deviation of $0.010$. 
The $p=1$ assumption will be taken later on just for economy.

The previous discussion about selection effects on the diagrams of Fig. \ref{LEvsT}
can be translated now into the $\Epr$ vs $\Liso \tdur^p$ correlation (Fig. \ref{LTESB}). 
We conclude then that the latter correlation is reasonably free of selection effects. 
This is rather evident if we imagine that the effect of \tdur\ is to shift the borders of the 
$\Epr$ vs $\Liso$ correlation into the body of the $\Epr$ vs $\Liso \tdur^p$ correlation. 
Then the lack of low luminosity events influences now the population of the 
$\Epr$ vs $\Liso \tdur^p$ correlation and not its borders.

\begin{figure*}
\vspace{11cm}
\hspace{13cm}
\includegraphics{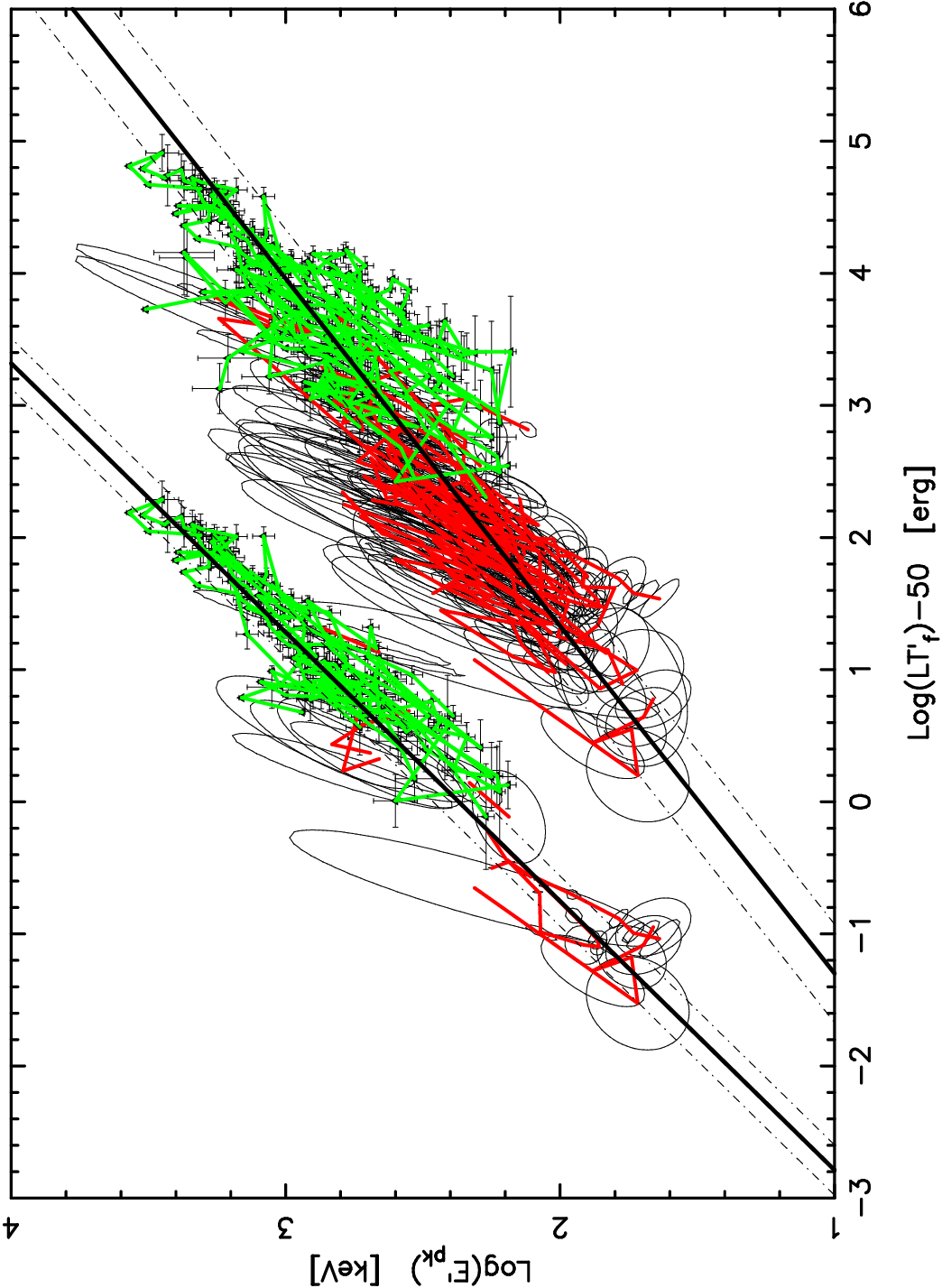}
\caption{
The data and lines in the right part of the diagram are the same as those presented in 
Fig. \ref{LESB} (the \Liso\tdur--\Epr\ diagram). The data in the left part correspond 
to the GRBs of both \sw\ and \cgro\ samples with known \tbr. For such GRB, \Lga\ has 
been used instead of \Liso\ (the HM case was used to calculate \thj). The symbol and 
line codes are the same as in Fig. \ref{LESB}. 
Note how the intrinsic scatter has been reduced  by going from the \Liso\tdur--\Epr\ 
diagram to the \Lga\tdur--\Epr\ one.
}
\label{LTCESB}
\end{figure*}

\begin{figure*}
\vspace{11cm}
\hspace{13cm}
\includegraphics{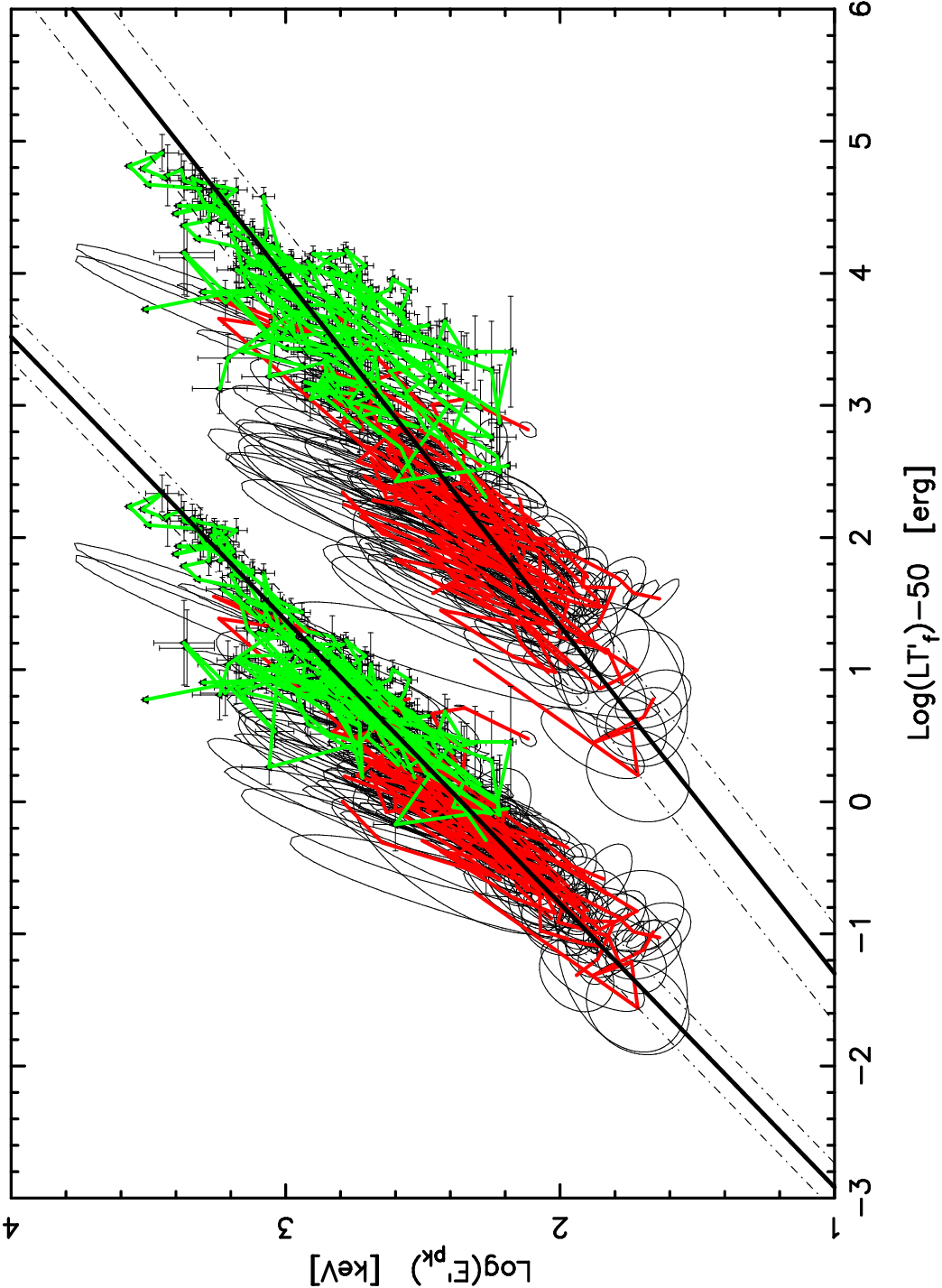}
\caption{The data and lines in the right part of the diagram are the same as presented in 
Fig. \ref{LTESB} (the \Liso\tdur--\Epr\ diagram). The data on the left correspond 
to the rigid shift of each evolutionary track to best fit with the $\Epr$ vs $\Lga\tdur$
correlation for the \sw\ sub--sample of GRBs with \tbr\ known (left part of Fig. 
\ref{LTCESB}). The \cgro\ data are shown only for illustrative purposes; 
Note that the scatter
of the $\Epr$ vs $\Lga\tdur$ correlation constructed by the track shifts is only
slightly smaller than  the one of the original collimation--corrected correlation.
}
\label{LTTESB}
\end{figure*}

\subsection{Scatter reduction by correcting luminosities due to collimation}

Following Ghirlanda et al. (2004), a way to reduce the scatter around the $\Epr$ 
vs $\Liso \tdur^p$ correlation could be by correcting the time--resolved isotropic 
luminosities by the jet collimation angle in order to estimate the intrinsic 
$\gamma-$ray temporal luminosities, $\Lga = (1 - \cos \thj)\Liso$. 
The collimation semi--aperture angle \thj\ is calculated from the jet break time \tbr\
and the total emitted isotropic energy \Eiso\ by two alternative models: 
the homogeneous ISM model (HM) and the wind medium model (WM).
Unfortunately, reliable estimates of \tbr\ for our GRB sample are available only for some 
events. 
We use the \tbr\ values compiled from the literature by Ghirlanda et al. 
(2007)\footnote{ 
Note that, in principle, \tbr\ should be achromatic, since it is due
to a geometrical effect. 
However, we rarely have true achromatic breaks in the well sampled 
light curves of \sw\ bursts. 
This may be due to the fact that the optical and the X--ray emission 
are due to two different components (see e.g. Uhm \& Beloborodov 2007; Genet, Daigne \&
Mochkovitch 2007; Ghisellini et al. 2007).  
As discussed in Ghirlanda et al. (2007), it is likely that the optical emission is 
more often associated to the forward shock of the fireball running into the 
circumburst medium, and therefore more indicative of possible jet breaks.}.
Our GRBs with known \tbr\ reduces to 7 events of \sw\ with 37 time resolved spectra, and 5 of 
\cgro\ with 195 time resolved spectra  (see Tables 1 and 2). 

In the HM model, the jet opening angle is given by:
\begin{equation}
\thj=0.161 {\left( {\frac{\tbr_{,d}}{1+z}} \right)}^{3/8}
\left( {\frac{n \eta_{\gamma}}{\Eiso_{,52}}} \right)^{1/8},  
\label{thetaH}
\end{equation}
where $\eta_{\gamma}$ is the radiative efficiency that we assume equal to 0.2, and $n$ 
is the medium density that we assume equal to 3 cm$^{-3}$.

Figure \ref{LTCESB} shows the collimation--corrected correlation 
$\Epr$ vs. $\Lga \tdur$ for 12 GRB with 
known \tbr\ (left correlation), together with the same data presented in Fig. \ref{LTESB} 
(right correlation). The best fit line showed for the collimation--corrected correlation refers 
only to the \sw\ sample; the corresponding scatter is represented by the dot--dashed lines.
Fitting the data with the line
\begin{equation}
{\rm Log}\left(\frac{\Epr}{\rm keV}\right) =  
a {\rm Log}\left( \frac{\Lga \tdur}{10^b{\rm erg}}\right)+c
\label{corrtg} 
\end{equation}
the best fit parameters of the $\Epr$ vs $\Lga \tdur$ correlations (HM case) for \sw\ and
\cgro\ samples are given in Tables 3 and 4, respectively (relation LTHE).
The slopes of both correlations are very close, the average being $0.50 \pm 0.03$.  
Concerning the scatter,  for both samples \sintr\ has clearly reduced its value with 
respect to the one in the $\Epr$ vs $\Liso \tdur$ correlations; on average we find
$\sintr = 0.10 \pm 0.01$.

We conclude that the HM collimation angle introduces a 
remarkable reduction on the scatter in the $\Epr$ vs $\Liso \tdur$ correlation.
However, such a reduction does not concern the internal scatter in each GRB evolutionary track.

In order to estimate the contribution of the internal scatter from each
evolutionary track to the overall correlation intrinsic scatter, we have
performed the following exercise. For each sample (\sw~ or \cgro), every \Liso\tdur--\Epr\ 
GRB track is shifted rigidly to best fit with the collimation--corrected correlation $\Epr$ vs 
$\Lga \tdur$ presented above for the corresponding sample 
(see Fig. \ref{LTTESB} for the \sw\ sample; here the evolutionary tracks of the \cgro\ 
GRBs are included only for graphic display).
We are now able to calculate the new intrinsic scatter of the corresponding sample, 
\sinto, around the respective correlation $\Epr$ vs $\Lga \tdur$. 
The values of \sinto\ (relation LTHE) are reported in Tables 3 and 4 for the \sw\ and \cgro\ 
samples, respectively.  
By comparing \sintr\ and \sinto, it is rather evident that the scatter of the tight
collimation--corrected correlation is dominated by the internal scatter of the
evolutionary tracks.  
A similar result is obtained if instead of the entire sample we estimate the \sinto\ taking into
account the GRB with the known jet break time alone. 
In fact in this case the average internal scatter is $\sinto=0.08 \pm 0.01$. 
We conclude that any further reduction of the
collimation--corrected correlation scatter could be reached by reducing the
internal scatter of each evolutionary track, and that the latter
should be possibly identifying the hidden parameters behind the stochastic
properties of the evolutionary tracks of each GRB.

In the WM model the jet opening angle is given by
\begin{equation}
\thj=0.202 {\left( {\frac{\tbr_{,d}}{1+z}} \right)}^{1/4}
\left( {\frac{A_{\ast} \eta_{\gamma}}{\Eiso_{,52}}} \right)^{1/4},
\label{thetaW}
\end{equation}
where now the medium density is supposed to be $n\left( r \right) = 5 \times
10^{11} A_{\ast} r^{-2}$ g cm$^{-1}$ and we assume $A_{\ast} = 1$.  Tables 3
and 4 present the corresponding best fit parameters for the $\Epr$ vs $\Lga$
\tdur\ correlation in the WM case (relation LTWE). In this case the best fit
slopes are moderately close ($95 \%$ CL). The average of both slopes is $0.60\pm 0.05$.  
The scatters \sintr\ of the WM collimation--corrected correlations
are also smaller than the ones of the not corrected correlations $\Epr$ vs $\Liso\tdur$, 
but the scatter reduction is smaller than in the HM case. On
average we find $\sintr = 0.12 \pm 0.01$.
After performing the same {\it track shifting} procedure described
above, we find that also in this case the internal scatter of each GRB
evolutionary track provides the dominant component of the scatter around the
tight $\Epr$ vs $\Lga\tbr$ correlation.

Finally, we have explored whether the residuals of the
\Liso\tdur--\Epr~ correlation correlate or not with several prompt light--curve
parameters: the variability $V$ (Reichart et al. 2001); the "emission symmetry" $S_F=T_2/T_1$,
where $T_1$ and $T_2$ are the duration times of the fluence--halves, from $5-50\%$ and 
$50-95\%$ of the total counts, respectively (Borgonovo \& Bj\"ornsson 2006); the
$\Ep^{(1)}/\Ep^{(2)}$ ratio, where $\Ep^{(1)}$ and $\Ep^{(2)}$ are the peak energies of the
integrated $\nu f_\nu$ spectra for each of the two time intervals $T_1$ and $T_2$, respectively 
(Borgonovo \& Bj\"ornsson 2006). Our preliminary results show that the residuals are not
correlated with any of these parameters, i.e. none of them could be a potential reductor of the 
scatter around the \Liso\tdur--\Epr~ correlation.

\section{Conclusions}

We have selected the high--signal time--resolved spectra from the available sample 
of \sw\ long GRBs with measured $z$, and analysed them with the aim to search for 
systematic features of the local $\gamma-$ray emission mechanism and their connection 
with known global GRB properties. Requiring a GRB peak flux $\ppf>1.8$ \funits,  
a total of 207 time--resolved spectra corresponding to 32 GRBs (until February 2008) 
were analysed. We have included also 287 spectra from 7 bright \cgro\ GRBs with $z$ 
known analysed previously in Kaneko et al. (2006). Since the two samples are 
affected in a different way by some systematic effects, we preferred to perform
our correlation analysis separately for each sample and then to check whether
the results are consistent or not between them. We have found that they are indeed 
consistent. Thus, for simplicity, in what follows we report the averages of the two 
samples for the best--fit parameters of the different correlations.

The main results and conclusions from our study are as follows:

\begin{itemize}

\item
By plotting the time--resolved data--points in the logarithmic \Liso--\Epr\ diagram, 
a linear band with average slope $0.30 \pm 0.05$ and intrinsic scatter $\sintr=0.21 \pm 0.01$ 
appears. While the low \Epr\ -- high \Liso\ region is free of selection effects, the high 
\Epr -- low \Liso~ region could be affected by the flux limits of the samples. 

\item 
We found that the residuals in the \Liso--\Epr\ diagram correlate with \tdur.
 This result offers a strong evidence that the parameter \tdur\ reduces the scatter of the 
$\Epr$ vs $\Liso$ correlation.   
By analyzing the \tdur\ vs $\Liso/\Epr^2$ diagram (Fig. \ref{LEvsT}), we have checked that selection 
effects are not responsible for such a trend. 

\item 
In agreement with the previous point, we have introduced the logarithmic diagram
$\Epr$ vs $\Liso \tdur^p$ and have found that the optimal value $p =1.1 \pm 0.1$ 
reduces the scatter to $\sintr=0.15 \pm 0.02$.
The average slope of the correlation is $0.40 \pm 0.02$.
Such correlation reveals three important aspects. First, its intrinsic 
scatter is smaller than the intrinsic scatter in the $\Epr$ vs $\Liso$ correlation. 
Second, it is reasonably free from selection effects, accordying to our our analysis
in the \tdur\ vs $\Liso/\Epr^2$ diagram. Third, it represents a 
connection between instantaneous features (\Epr, \Liso) and global features (\tdur).
{\it At any moment, the instantaneous features of a GRB correlate with the entire 
prompt duration as if at each instant of the prompt the GRB {\it would be aware} 
of the duration of the entire process}. 

\item 
For the 12 GRBs out of our samples (7 from the \sw\ and 5 from the \cgro\ samples, 
respectively) for which the jet break time \tbr\ is known, we could further reduce the scatter 
of each sample by using the collimation--corrected luminosity \Lga\ by estimating the 
collimation jet angle for the homogeneous (HM) and wind medium (WM) cases. 
The lowest intrinsic scatter has been obtained for the $\Epr$ vs $\Lga \tdur$ correlation 
in the HM case; the (average) slope of the correlation is $0.50 \pm 0.03$ 
and $\sintr = 0.10 \pm 0.01$.

\item 
We have estimated the contribution of the internal scatter of the evolutionary
tracks to the scatter of the overall correlations. For this, the \Liso--\Epr\
evolutionary track of each GRB has been shifted to best fit with the corresponding
collimation--corrected (HM and WM cases) correlations.
Our results indicate that for both cases $\sintr \approx \sinto$,
this means that the total intrinsic scatter is mainly due to the internal scatter of the tracks.
Thus, with the caveat that the statistics is still limited, we conclude that 
{\it any further reduction of the scatter around the GRB empirical correlations may be 
attained by discovering the hidden variables behind the stochastic features of the 
individual \Epr--Liso\ evolutionary tracks}.

\end{itemize}

We conclude that the long GRB individual evolutionary tracks
populate a rather narrow strip in the $\Epr$ vs $\Liso \tdur$ diagram with a slope 
$\approx 0.4$, whatever the evolutionary track slope is. 
The jet collimation correction further reduces the thickness of such strip and
leads the slope close to $0.5$.
While selection effects probably are present in the $\Epr$ vs $\Liso$ diagram, 
they do not seem to weaken our conclusion.
This implies the existence of a universal $\gamma$--ray emission mechanism for long 
GRBs where the instantaneous features are modulated by a global parameter, which 
we found here to be \tdur.  
We suggest that the interconnection between \tdur\ and the \Liso--\Epr\ evolutionary tracks 
is at the basis of the global `Amati' and `Ghirlanda' relations.

\section*{Acknowledgments}

We thank PAPIIT--UNAM grant IN107706 to V.A. and the italian INAF and 
MIUR (Cofin grant 2003020775\_002) for funding.

\end{document}